# Band Structure, Phonon Scattering and the Ultimate Performance of Single-Walled Carbon Nanotube Transistors

Xinjian Zhou[1], Ji-Yong Park[2], Shaoming Huang[3], Jie Liu[3], and Paul L. McEuen[1]
[1] Laboratory of Atomic and Solid-State Physics, Cornell University, Ithaca, New York 14853 USA
[2] Department of Physics, Ajou University, Suwon 442-749 Korea
[3] Chemistry Department, Duke University, Durham, North Carolina 27708 USA

Semiconducting single-walled carbon nanotubes are studied in the diffusive transport regime. The peak mobility is found to scale with the square of the nanotube diameter and inversely with temperature. The maximum conductance, corrected for the contacts, is linear in the diameter and inverse temperature. These results are in good agreement with theoretical predictions for acoustic phonon scattering in combination with the unusual band structure of nanotubes. These measurements set the upper bound for the performance of nanotube transistors operating in the diffusive regime.

Semiconducting single walled carbon nanotubes (SWNTs) are remarkably high-performance electronic materials. The SWNT band structure, shown in Fig. 1(a), is that of relativistic one-dimensional fermions, with energies and velocities given by:

$$E = \pm\sqrt{(m^* v_0^2)^2 + (\hbar k v_0)^2} \; ; \; v = \frac{1}{\hbar}\frac{dE}{dk} = \frac{\hbar v_0^2 k}{E} \quad (1)$$

where $v_0 = 8 \times 10^5$ m/s, the Fermi velocity in graphene, plays the role of the speed of light. The carrier effective mass is set by the tube diameter $d$: $m^* = \frac{2\hbar}{3 d v_0}$ [1]. In the absence of disorder, scattering of electrons will occur due to the dynamical fluctuations in the tube itself, i.e. phonons. The scattering rate $\tau^{-1}$ should scale as the number of phonons available for scattering, i.e. proportional to $T$. This phonon scattering, coupled with the unusual band structure, should set the ultimate performance limits of SWNT transistors.

Field-effect transistors (FETs) were first made from SWNTs several years ago [2] and have been subsequently investigated intensely for device [3-6] and sensing [7, 8] applications. Both Schottky [9] and low resistance [10, 11] contacts have been realized, and short devices have been shown to operate at near the ballistic limit [10]. However, the ultimate performance limits are not understood; for example, the reported mobility values vary by orders of magnitude in different studies [6, 12-14]. Here we systematically study the properties of moderately long (4-15 $\mu$m), oriented SWNT FETs with good contacts to probe the intrinsic transport properties. We find that both the temperature and diameter dependence of the mobility and maximum conductance are well described by the predictions of acoustic phonon scattering in concert with the "relativistic" band structure of nanotubes.

The devices used in this experiment are shown schematically in Fig. 1(d). Using the recipe in Refs. [15] and [16], SWNTs are grown on silicon wafers with 200 nm oxide. Figure 1(c) is an atomic force microscope (AFM) image of one of the devices. The tubes are long with little mechanical bending and oriented with the gas flow direction in the CVD growth furnace. The electrodes are made of either Au or Pd to make ohmic contact. In order to keep the SWNT free from rare defects, we choose the channel lengths $L$ to be from 4 $\mu$m to 15 $\mu$m.

Figures 2 and 3 show the low-bias conductance in the p-type region for three SWNT FETs with different diameters as a function of backgate voltage [17]. The main panel of Fig. 2 shows the temperature dependence of the conductance of one device with $d = 4$ nm and $L = 4$ $\mu$m. The transition from off to on is more rapid and the maximum conductance increases with decreasing temperature. The resistance of the on-state, which is defined as when $V_g$ is 10 V away from threshold voltage, decreases linearly with decreasing temperature down to 50 K, as shown in the upper inset, with an extrapolated intercept at zero temperature of *28 kΩ*.

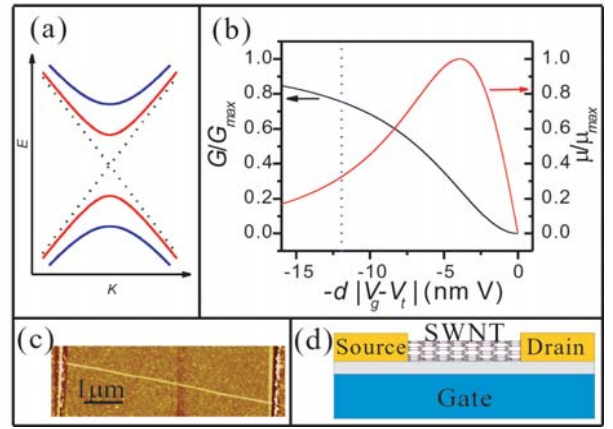

FIG. 1. (color online). (a) The relativistic band structure of semiconducting SWNT. (b) Theoretical plot of SWNT FET conductance and mobility as functions of gate voltage according to Eqs. (5) and (6). The dotted line indicates where the Fermi level reaches the second subband at zero temperature. (c) AFM image of one device. (d) Schematic of device geometry. The gray layer is 200 nm thick silicon oxide between SWNT and highly-doped silicon wafer used as backgate.

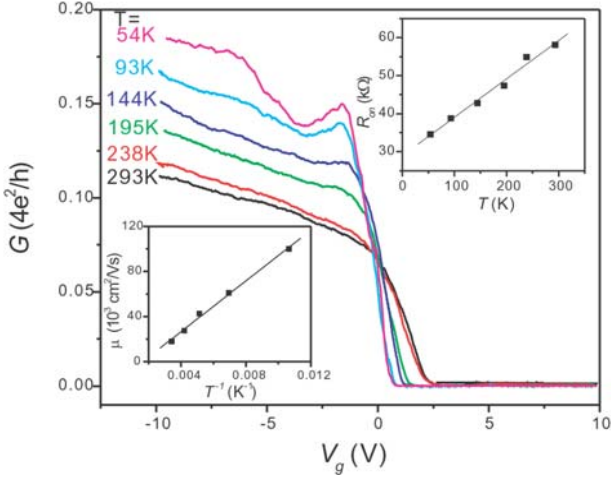

FIG. 2. (color online). The conductance of a device with $d = 4$ nm and $L = 4$ μm versus gate voltage at different temperatures. Upper inset: $R_{on}$ as a function of $T$. Lower inset: measured mobility as a function of $T^{-1}$. Both are shown with linear fitting.

Figure 3 shows data from two other semiconducting SWNT FETs of different diameters. The maximum on-state conductance increases with $d$; larger tubes have larger on-state conductance, and their conductance increases more dramatically upon cooling. This is shown in Fig. 4(a), where data from a number of devices of different diameters are compiled. To minimize the influence of contact resistance, the temperature-dependent change is measured and is found to approximately scale linearly with tube diameter $d$.

Shown in Fig. 3(b) are the field-effect mobilities for the two devices in Fig. 3(a) calculated using $\mu_{FE} = (L/C_g')(dG/dV_g)$, where $C_g'$ is the gate capacitance per unit length. $C_g'$ is estimated to be $2 \times 10^{-11}$ F/m [6, 13] for $d = 4$ nm tube and it scales with diameter according to $C_g'^{-1} \propto \ln(1+2\lambda/d)$ [6], where $\lambda = 200$ nm is the oxide thickness. The peak mobility is approximately 10,000 cm$^2$/V s for the $d = 3.4$ nm, $L = 5.4$ μm tube, but about four times lower for the $d = 1.5$ nm, $L = 10$ μm tube. This is characteristic of all the devices measured; smaller-diameter tubes have lower peak mobilities, as is shown in Fig. 4(b). With the exception of a few devices with anomalously high mobilities (marked by circles in the plot), the data vary approximately as the square of the diameter (see inset to Fig. 4(b)).

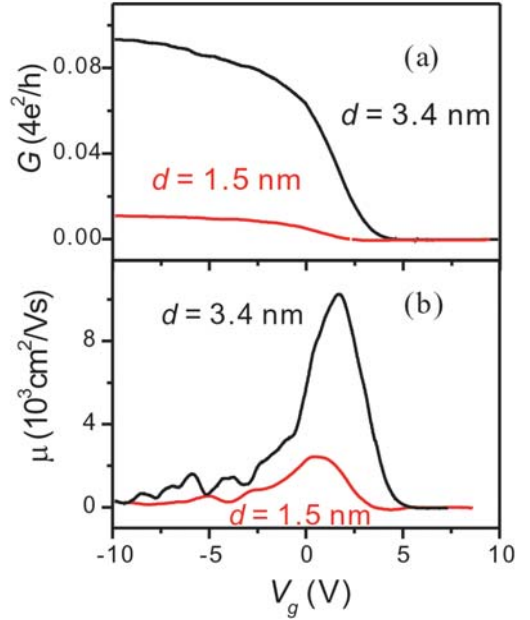

FIG. 3. (color online). The conductance (a) and mobility (b) as functions of gate voltage for two different devices. The black curves are for a tube with $d = 3.4$ nm, $L = 5.4$ μm while the red curves are for a tube with $d = 1.5$ nm, $L = 10$ μm.

To analyze these results, we first note that the device resistance is a combination of the intrinsic tube resistance $R_{NT}$ and contact resistance $R_c$. To separately determine these contributions, we used an AFM tip as an electrical nanoprobe [11]. For the device in Fig. 2, at room temperature we find: $R_c \approx 28$ kΩ and $R_{NT} \approx 30$ kΩ at on-state. Note that $R_c$ is the same as the extrapolated tube resistance at zero temperature (Fig. 2, upper inset). We therefore conclude that $R_c$ is temperature-independent, and it is $R_{on}$, which is $R_{NT}$ at on-state, that varies linearly with $T$, extrapolating to zero at $T = 0$. At room temperature, $R_{on}/L$ is about 8 kΩ/μm for this semiconducting tube. This value is comparable to but somewhat higher than the resistivity of good metallic tubes [19, 20] and is similar to reported values for semiconducting tubes [13, 14]. At the lowest temperature shown ($T = 50$K), the device resistance is small ($R_{on} < 10$ kΩ), but still finite, corresponding to a minimum resistivity $R_{on}/L \sim 2.5$ kΩ/μm. At lower temperatures, the device resistance begins to increase (not shown), presumably due to the effects of localization and/or Coulomb blockade [20, 21].

The lower inset to Fig. 2 shows the peak mobility of the device at different temperatures after the constant contact resistance is taken into account. It varies approximately as $T^{-1}$, reaching $> 100,000$ cm$^2$/V s at 50 K. To summarize the experimental findings: $R_{on}$ and $\mu^{-1}$ both increase linearly with $T$. Both depend strongly on diameter, but in different ways: $1/R_{on} \sim d$ and $\mu \sim d^2$.

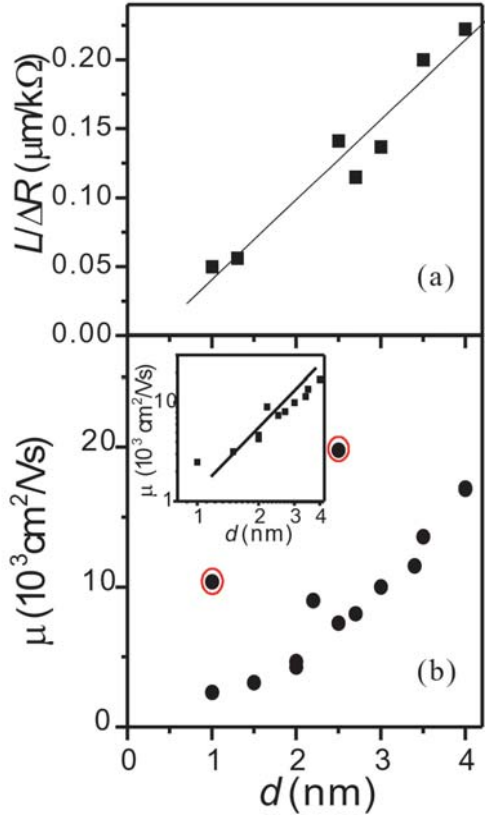

FIG. 4. (color online). (a) Device length $L$ divided by resistance change $\Delta R$ from room temperature to 50 K versus tube diameter with linear fitting shown. Data of similar diameter tubes are averaged. (b) Room temperature peak mobility as a function of tube diameter for many devices. Two devices showing anomalously large peak mobility are circled. Inset to (b): peak mobilities versus tube diameters on log-log scale with circled two data points omitted. The solid line is a visual reference line for square law.

To understand these results, we start with the Drude model in 1D conductor with four channels [22]:

$$G_{NT} = \frac{4e^2}{h}\frac{\tau_F v_F}{L} = \frac{4e^2}{h}\frac{\ell_F}{L} \qquad (2)$$

where $\ell_F = \tau_F v_F$ is the mean free path at the Fermi level. In 1D, the Fermi wave vector is a linear function of the gate voltage away from threshold $\Delta V_g = |V_g - V_t|$:

$$k_F = \frac{\pi n}{4} = \frac{\pi C_g' \Delta V_g}{4e} \qquad (3)$$

By Fermi's Golden Rule, the scattering rate is proportional to the density of states, which in 1D scales as $1/v$:

$$\tau^{-1} = \tau_0^{-1}(v_0/v) \qquad (4)$$

where $\tau_0^{-1}$ is the scattering rate at high velocities.

Combining Eqs. (1) - (4) gives the following expression for the conductance:

$$G(V_g) = \frac{4e^2}{h}\frac{\ell_0}{L}\frac{(\Delta V_g/a)^2}{1+(\Delta V_g/a)^2}$$

where $\quad a = \frac{8e}{3\pi d C_g'} \qquad (5)$

Here $\ell_0 = v_0 \tau_0$ is the mean free path at high energies. This expression is plotted in Fig. 1(b). The conductance rises over a gate voltage range determined by the parameter $a$, then saturates as $v_F$ approaches $v_0$. The field-effect mobility is calculated to be:

$$\mu_{FE} = \frac{e\tau_0}{m^*}\frac{(\Delta V_g/a)}{\left(1+(\Delta V_g/a)^2\right)^2} \qquad (6)$$

The mobility rises linearly with $\Delta V_g$ before peaking and drops back toward zero as $v_F$ approaches $v_0$, as shown in Fig. 1(b). The peak value is:

$$\mu_{peak} = 0.32\frac{e\tau_0}{m^*} \qquad (7)$$

The Fermi energy reaches the second subband at large gate voltage as the dotted line in Fig. 1(b) indicates. However, the inter-subband scattering makes the addition of more conduction channels less noticeable so that the resistance does not deviate much from our simple model [23]. Plus the electrons in higher subbands experience large Schottky barrier when they get into SWNTs and the conductance through these channels is lowered [24]. So in the theoretical analysis above, we can take only the first subband into consideration while neglecting the influence of higher subbands.

The theory graph Fig. 1(b) qualitatively describes the data in Figs. 2 and 3. To make a quantitative comparison, we use the following expression for the scattering rate from acoustic phonons [23, 25, 26]:

$$\tau_0^{-1} = \alpha\frac{T}{d} \qquad (8)$$

Combining Eqs. (5), (7), and (8) gives the following results for the peak mobility and maximum conductance:

$$\mu_{peak} = 0.48\frac{ev_0}{\hbar\alpha}\frac{d^2}{T}; \quad G_{max} = \frac{4e^2 v_0}{h\alpha L}\frac{d}{T} \qquad (9)$$

These expressions correctly predict the temperature and diameter dependences observed in Figs. 2-4. From the data in Fig. 2 we can extract values for the coefficient α from either the maximum conductance or the peak mobility: α = 9.2 m/K s and α = 12 m/K s.

Reference [26] obtained a value of $\alpha \approx 6.5$ m/K s for the total scattering rate. More recently, using Monte Carlo simulation, Ref. [27] obtained nearly identical function forms to Eq. (9) but with a smaller value $\alpha \approx 1.6$ m/K s. The origin of the discrepancy is not known.

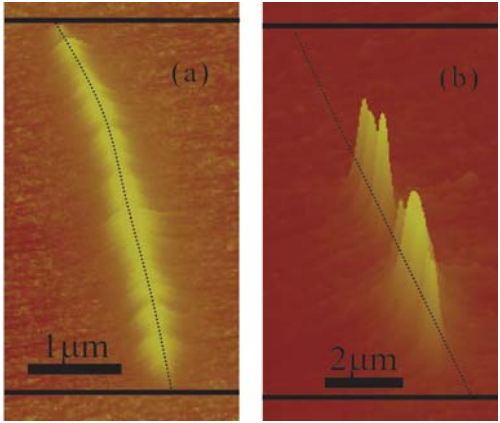

FIG. 5. (color online). Scan gate microscopy images of (a) the device studied in Fig. 2 with normal peak mobility and (b) the circled $d$ = 2.5 nm device with anomalously high peak mobility in Fig. 4(b). The solid lines indicate the position of electrodes and dashed lines indicate the position of SWNT. The device in (b) is dominated by a few short segments of tube near the threshold region. Therefore the mobility is overstated when the whole tube length is used to calculate the mobility.

A few of the devices showed anomalously large peak mobilities, as indicated by the circled data points in Fig. 4(b). To understand the origin of this behavior, we used scanned gate microscopy [28] to explore the uniformity of the conduction near the threshold region. In brief, a biased AFM tip is used as a local gate and the conductance of the device is recorded as a function of the AFM tip position. The results are shown in Fig. 5. For most devices, the response to the tip was relatively uniform [Fig. 5(a)], indicating that no single spot dominated transport. For the devices with anomalously high mobilities, we found that the gate response was primarily localized to a few dominant spots. In this case, using the equation $\mu_{FE} = (L/C_g^{'})(dG/dV_g)$, where $L$ is the entire length of the tube, dramatically overstates the true mobility. This effect may also explain some of the extremely high values for the mobility reported in the literature [13].

In conclusion, SWNT FETs were studied in the diffusive regime. The on-state conductance increases linearly with the tube diameter $d$, and the peak mobility grows as $d^2$. Both quantities grow linearly with $T^{-1}$. These observations are in good agreement with the predictions for scattering by low-energy phonons coupled with the 1D "relativistic" band structure. These results set the ultimate performance limits of SWNT transistors operating in the diffusive regime.

This work was supported by the NSF Center for Nanoscale Systems, by the MARCO Focused Research Center on Materials, Structures, and Devices and by the Nanobiotechnology Center (NBTC), an STC Program of the National Science Foundation under Agreement No. ECS-9876771. Sample fabrication was performed at the Cornell node of the National Nanofabrication Users Network, funded by NSF.